\documentclass[runningheads]{llncs}
\usepackage{graphicx}
\usepackage{cite}
\usepackage{amsmath,amssymb,amsfonts}
\usepackage{algorithmic}
\usepackage{textcomp}
\usepackage{xcolor}
\usepackage{booktabs}
\usepackage{comment}

\graphicspath{{./figures/}}

\begin{document}
\title{POKs Based Secure and Energy-Efficient Access Control for Implantable Medical Devices}
\titlerunning{POKs-based Access Control for Implementable Medical Devices}
%
\author{Chenglong Fu\inst{1}\and
	Xiaojiang Du\inst{1} \and
	Longfei Wu\inst{2}\and
	Qiang Zeng\inst{3}\and
	Amr Mohamed\inst{4}\and
	Mohsen Guizani\inst{4}
}
\institute{Temple University, Philadelphia PA 19122, USA
	\email{\{chenglong.fu,xjdu\}@temple.edu} \\ \and
	Fayetteville State University, Fayetteville NC 28301, USA
	\email{lwu@uncfsu.edu}\\ \and 
	University of South Carolina, Columbia SC 29208, USA
	\email{zeng1@cse.sc.edu} \\ \and
	Qatar University, Doha, Qatar
	\email{\{amrm,mguizani\}@ieee.org}
}

\authorrunning{C. Fu et al.}

\maketitle              

\begin{abstract}
Implantable medical devices (IMDs), such as pacemakers, implanted cardiac defibrillators and  neurostimulators are medical devices implanted into patients' bodies for monitoring physiological signals and performing medical treatments. Many IMDs have built-in wireless communication modules to facilitate data collecting and device reprogramming by external programmers. The wireless communication brings significant conveniences for advanced applications such as real-time and remote monitoring but also introduces the risk of unauthorized wireless access. The absence of effective access control mechanisms exposes patients' life to cyber attacks. In this paper, we present a lightweight and universally applicable access control system for IMDs. By leveraging Physically Obfuscated Keys (POKs) as the hardware root of trust, provable security is achieved based on standard cryptographic primitives while attaining high energy efficiency. In addition, barrier-free IMD access under emergent situations is realized by utilizing the patient's biometrical information. We evaluate our proposed scheme through extensive security analysis and a prototype implementation, which demonstrate our work's superiority on security and energy efficiency.


\end{abstract}

\section{Introduction}
\textit{Implantable medical devices} (IMDs) are electronic devices that can be either partially or fully implanted into patients' bodies for collecting patients' physiological data and delivering timely treatment. With advantages of providing ongoing diagnosis and treatment, IMDs have been pervasively used for treating chronic medical disorders and are proved to be effective in coping with sudden deaths caused by cardiac arrests and ventricular arrhythmia \cite{pope2001innovation}. Most IMDs are equipped with radio modules for wireless communication with external devices called \textit{programmers} which are used by physicians for data exporting and IMD reprogramming. However,  the wireless communication brings attackers extra arsenal to threaten the users' lives with cyber attacks. Jay Radcliffe and Branaby Jack have demonstrated the feasibility to remotely hack the insulin pump in~\cite{radcliffe2011hacking} and~\cite{jack2013implantable} respectively. Coincidentally, former U.S. Vice President Dick Cheney had the wireless connection of his implanted defibrillator disabled due to the concern of cyber attacks launched by terrorists \cite{DickCheney}.

Although countermeasures against unauthorized IMD access are critical, the design is challenging due to two technical difficulties. First, requirements for utility and security are conflicting. On the one hand, the scheme must be robust enough to defeat all malicious access. On the other hand, during the emergent situation, first-aiders may need to access or reprogram a patient's IMD immediately without any hindering caused by security mechanisms. Second, the IMD security mechanism must be extremely energy efficient because IMDs are implanted into human's bodies via surgeries and rely on their embedded batteries to operate many years.

To cope with these security problems of medical devices, a series of research works are proposed \cite{hei2010defending,hei2013pipac,xia2017medshare}. Recent researches tend to address the first difficulty with the \textit{touch-to-access} principle which is based on a reasonable assumption that attackers having physical contact with victims can harm them directly (rather than utilizing IMDs). Following this, many proposed solutions \cite{xu2011imdguard,CCS2013,Gollakota2011} implement simple access control policies by verifying physical access or proximity for IMD access attempts. However, these proposed works have three drawbacks. (1) Their enforcement of the touch-to-access policy is \emph{not} based on provable security and may be breached by newly-developed attack techniques \cite{DAC2013flaw}.  (2) They assume IMDs are equipped with special sensing or communication capabilities like ECG measurement and piezo broadcasting. (3) Simple touch-to-access access control policy cannot deal with complicated scenarios like hierarchical privileges.

Our goal is to design an IMD access control solution that provides provable security without high energy consumption. To that end, we present a \emph{Physical Obfuscation Keys} (POKs) based IMD access control system. Leveraging a POKs enabled IC card for secure credential storage, we design a lightweight access control protocol with minimal computation and communication overhead on IMDs. For emergent access, We follow the touch-to-access principle and verify physical contact by requiring the patient's IC card and iris image. Our design is built on standard cryptographic operations to provide provable security and does not need any special sensing or communication capability of the IMD. Moreover, an online \emph{Hospital Authentication Server} (HAS) is integrated in our system to authenticate the programmer's identity and realize the dynamic and fine-grained access control.

In summary, our work has the following contributions:
\begin{itemize}
	\item  \textit{A novel POKs based key agreement scheme:} We innovatively use the POKs enabled IC card to design a secure key agreement scheme for IMDs. In our proposed protocol, computation-intensive operations are offloaded to the IC card and the Hospital Authentication Server to reduce the energy consumption of the IMD. 
	\item  \textit{Biometrics based emergent access:} We design a highly secure method that uses the patient's iris for barrier-free emergent access.
	\item  \textit{Real-device implementation:} We implement the IMD's logic on the TelosB sensor mote, analyze the security properties of the design, and evaluate the energy consumption to show our design's advantage of energy efficiency as well as its speed and memory consumption.
\end{itemize}

\par
The rest of the paper is organized as follows. We first describe some background on POKs and our key generation in Section \ref{sec:BF}. Then we present the system and threat model in section \ref{sec:model}. Our access control scheme for normal access and emergent access is described in Sections 
\ref{sec:protocol} and \ref{sec:enh}, respectively. After that, we analyze the security of our work in Section \ref{sec:eva} and the results of the overhead consumption evaluation are presented in Section \ref{sec:exp}. The review of the related works is in Section \ref{sec:related}. Finally, we conclude in Section \ref{sec:con}.

\section{Background and Key Generation}\label{sec:BF}
\subsection{Background: Physical Obfuscation Keys}\label{subsec:poks}
Modern crytographic primitives have their security based on the confidentiality of secret credentials. Once the credentials are uncovered, attackers can maliciously impersonate legitimate users or retrieve sensitive information from encrypted communications. Thus, secure secret key storage components such are critical to the security of all kinds of applications. However, recently developed physical tampering attacks such as micro-probing attack and electrical glitching attack have already showed their effectiveness on retrieving secret credentials kept in statical storage devices.

To defeat these physical attacks targeting secret credentials, researches propose the Physical Unclonable Functions (PUFs)~\cite{ravikanth2001physical} as a hardware-based cryptographic component for authentication and secret key storage. PUFs rely on unique physical characteristic variation as the secret challenge-response pattern. When given an input, the PUF responses with a corresponding unpredictable output. These patterns are determined by the PUF's unique and unclonable physical properties such as the integrated circuit's gate propagation delay\cite{lim2005extracting} and SRAM cell initial status\cite{holcomb2009power} which are introduced by uncontrollable variations during the IC fabrication process. Generally, PUFs are divided into two broad categories: strong PUFs and weak PUFs, differing in the number of available Challenge-Response Pairs (CRPs). Typically, strong PUFs can have a large set of CRPs while weak PUFs only support a limited number of CRPs. The Physically Obfuscated Keys (POKs) \cite{gassend2003physical} is an application of weak PUFs that could be considered as a secure key storage techinique. The appearance of this technique brings a reliable solution to protect the secret key from being compromised by invasive side-channel attacks. Recent POKs-based secure key generators have reached a bit error rate of less than 1\% \cite{karpinskyy20168} and require no special processing during the chip fabrication, making it a cost-efficient alternative to expensive secure storage components like EEPROM. 

\par
Bringer, \emph{et al}.\ make a further step by proposing POKs compatible cryptographic algorithms\cite{bringer2009physical} to defeat the runtime \textbf{memory scanning} attacks which aim to steal the credentials after they are loaded to the memory. They achieve it by splitting scalar product and exclusive OR (XOR), two basic operations used in the linear-feedback shift register (LFSR), into multiple steps. During each step, only part of the secret stored in the POKs are loaded to the memory with all intermediate results \emph{obfuscated}. Based on the specially designed LFSR, secure POKs enabled stream ciphers like Trivium~\cite{de2006trivium} is realized. Therefore, compared with the traditional primitives, POKs enabled cryptographic operations are resilient to memory scanning attacks.  
Moreover, 
\textbf{invasive tampering attacks} that try to unpack the chip will change physical features the keys rely on and thus destroy the the secret inside it permanently. Thus, revealing keys via invasive tampering attacks also fail. 

\subsection{Key Pair Generator}\label{sec:gen}

\par
\begin{figure}[ht]
    \centering
    \vspace{-5pt}
    \includegraphics[width=0.6\textwidth]{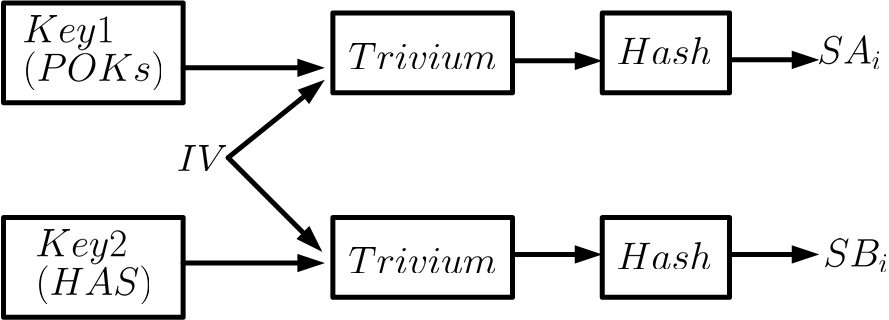}
    \vspace{-5pt}
    \caption{The key pair generator. $SA_i$ and $SB_i$ are generated by patient's IC card and the hospital authentication server respectively.}\label{fig:gen}
    \vspace{-10pt}
\end{figure}

For multi-party communication, key management (e.g., \cite{du2009transactions,xiao2007survey,du2007routing,du2007effective} is important for medical and IoT devices. We design a key generator to synchronously generate temporary keys for each access. As shown in Figure~\ref{fig:gen}, the key generator has the input of a master key and an initial value fed into a Trivium stream generator. During each access, 256 bits are truncated from the output stream and then be processed by the SHA-256 module to get a 256-bit temporary key. In the system, we have two different long-term master keys $Key_1$ and $Key_2$ to derive two temporary keys $SA_i$ and $SB_i$. $Key_1$ and $Key_2$ are stored by the patient's IC card and the HAS (Hospital Authentication Server), respectively, and the IMD stores the two master keys.

\begin{figure}[ht]
    \centering
    \vspace{-5pt}
    \includegraphics[width=0.5\textwidth]{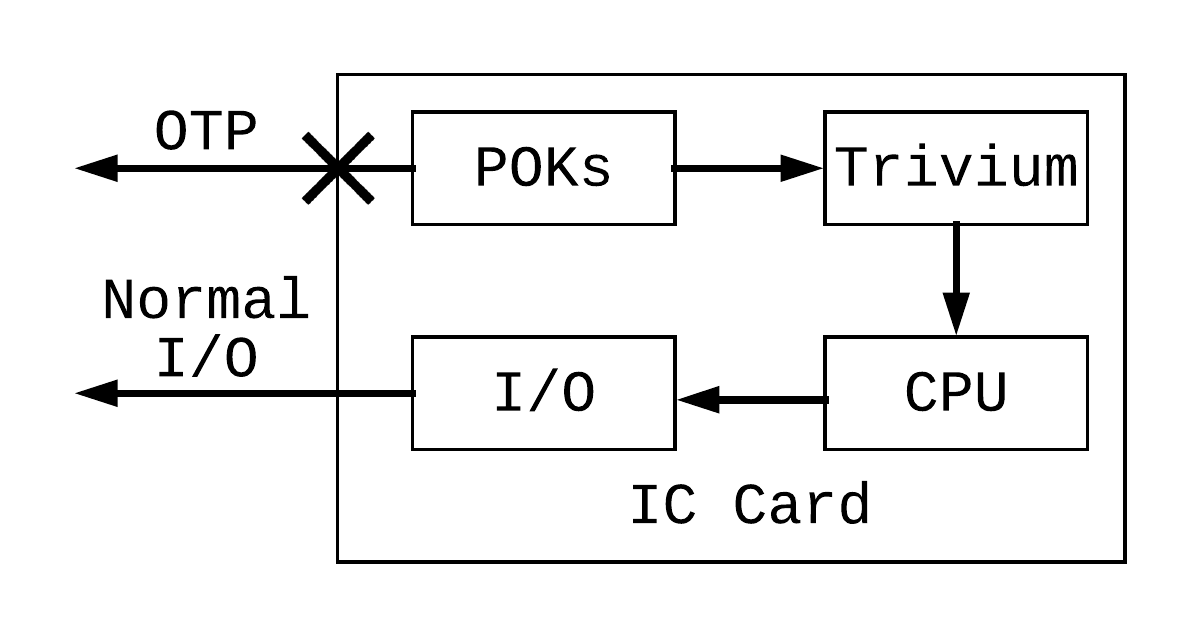}
    \vspace{-10pt}
    \caption{IC card structure.}\label{fig:POK_IC}
    \vspace{-5pt}
\end{figure}

On the IC card, we have a POK module to store the master key $Key_1$ and a POKs-based Trivium generator to generate temporary keys. As illustrated in the Figure~\ref{fig:POK_IC}, the IC card has the POK module, the Trivium stream generator, and the CPU packed on one chip. The secret stored by the POK module can only be accessed by the Trivium generator which outputs the bitstream directly to the embedded CPU for cryptographic operations. The One-Time Programming (OTP) interface \cite{min2005embedded} is added to enable one-time access to the secret during the IC card commissioning. After the first access, the OTP interface is physically disabled to prevent any direct access to the master key.

\par
The doctor's IC card also has a POK module to store the master key $Key_3$ that is used for generating 256-bit message authentication code (MAC). The MAC algorithm is based on LFSR as described in \cite{bringer2009physical}, which is inherently resistant to memory probe attacks and invasive attacks.

\section{System Overview and Threat Model} \label{sec:model}

\begin{figure}[h]
	\centering
	\includegraphics[width=0.6\textwidth]{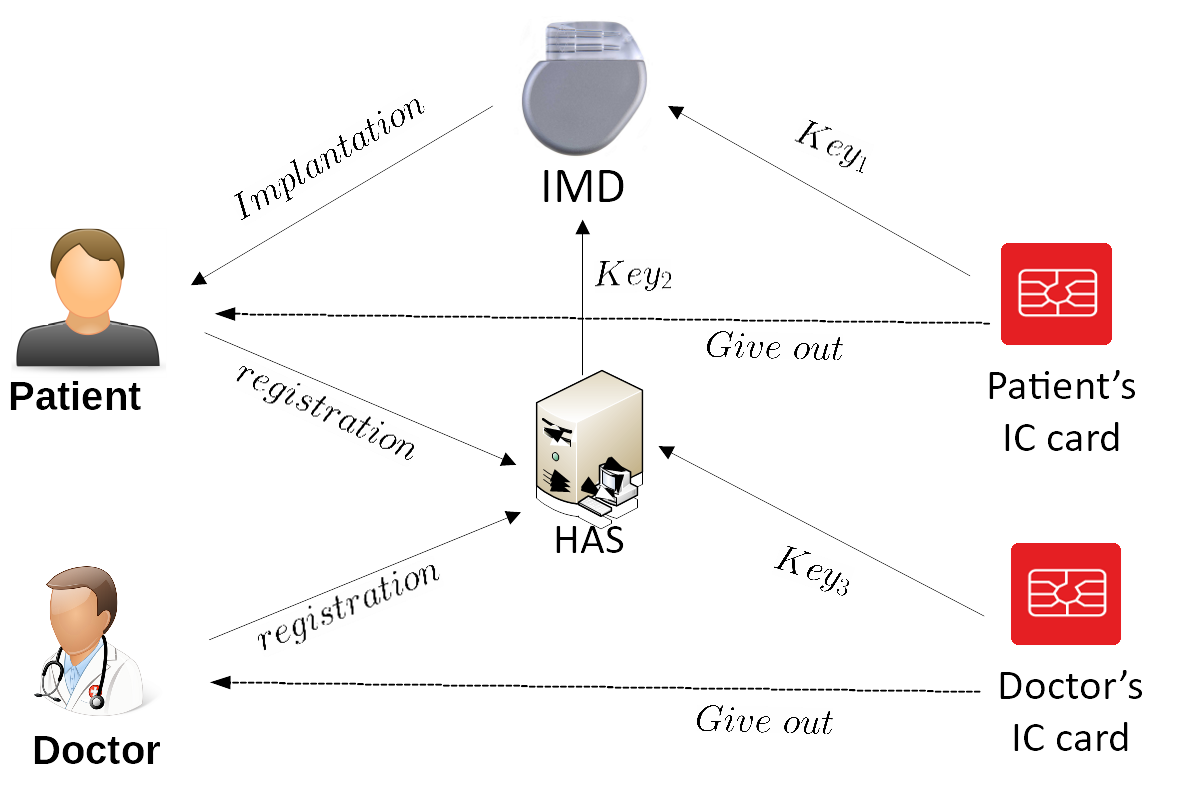}\\
	\caption{The enrollment phase.}\label{fig:enroll}
	\vspace{-5pt}
\end{figure}

\subsection{System Overview}
As shown in Figure~\ref{fig:enroll}, our access control system involves the following components: the IMD, the programmer, the Hospital Authentication Server (HAS), and IC cards for patients and doctors. IC cards are contact cards with embedded POKs modules. The HAS is a centralized authentication server that is expected to be common in modern hospital and e-health systems. Two pre-defined master keys are shared by the patient's IC card and the HAS with the IMD respectively. Each master key is used as the secret seed to generate different temporary keys for each access. Then the programmer authentication and the session key generation are all conducted with the temporary keys.

In our design, the patient's and doctor's IC cards are commissioned as the representation of their identities via the enrollment phase. During the regular access, the doctor firstly plugs his IC card into the programmer to prove his identity to the HAS. Then, the programmer forwards the challenge message received from the IMD to the IC card and the HAS. Finally, the programmer generates the correct response message with the help of the IC card and the HAS and establishes a secure communication channel with the IMD. The HAS enforces fine-grain access control policies according to the doctor's identity and the access operation type.

\subsection{Assumptions}
For devices, we assume the IMD and IC card are capable of running basic cryptographic operations including SHA-256 and HMAC. While the programmer has Internet access for communicating with the Hospital Authentication Server (HAS). The IC cards are contact cards without wireless communication capability. We can safely assume IC cards are taken with patients and physicians all the time considering its small size easy maintenance (no charging required).

As for the emergent situation when no Internet connection and valid physician's IC card are available, we assume first-aiders can find the patient's IC card have the equipment to acquire the patient's iris code.

\subsection{Threat Model}
In this paper, we assume a powerful adversary with abilities to eavesdrop all wireless communications between the programmer and the IMD and send arbitrary messages to the IMD remotely.

Also, we reasonably assume the adversary cannot replicate the IC card or retrieve the secret from it due to the POKs's unclonable feature. The secret keys stored in the IMD is also inaccessible because the IMDs are implanted into patients' body and are physically inaccessible. Also, the HAS is considered as secure because the HAS has plenty of resources for enforcing security schemes and is managed by professional security administrators.

\section{The Authentication Protocol}\label{sec:protocol}

\par
In this section, we present our IMD authentication and access control protocol, which is based on the a pair of temporary keys $SA_i$ and $SB_i$ and the doctor's master key $Key_3$. The protocol is composed of four phases: Enrollment, Service Request, Authorization, and Session Establishment. The following parties are involved: an Implantable Medical Device (IMD), a Hospital authentication Server (HAS), a Programmer (with the doctor's IC card plugged in), and the patient's IC card. Table~\ref{tab:note} summarizes all the symbols and notations used in our description. Figure~\ref{fig:enroll} and Figure~\ref{fig:flow} illustrate  the enrollment phase and all the following phases, respectively.

\begin{table}
	\scriptsize
	\centering
	\caption{Symbols and Notations.}
	\begin{tabular}{lp{4.0cm}}
		\toprule
		Notation & Description\\
		\midrule
        $Key3$ & doctor's master key\\
		$SA_i,\ SB_i$ & 256-bits  temporary key pair \\
		$R$ & 32-bits Service Request Code \\ 
		$i$ &32-bits counter\\ 
		$ID_I,\ ID_P,$ & 32-bits identity for patient and doctor \\
		$T_1$ & 32-bits Time stamp\\
		$TS$ & 32-bits Time window threshold\\ 
		$T$ & current time\\ 
		$token(i)$ & token for the ith cycle\\ 
		$S_{key}(i)$ & Session key for the ith cycle\\ 

		\bottomrule
	\end{tabular}
	\label{tab:note}
\end{table}

\subsection{Enrollment}
\par
During the enrollment phase, all parties in our protocol are initialized through two independent steps as shown in Figure~\ref{fig:enroll}.

\subsubsection{Doctor's Registration}
Since we assume the programmer is not bound with the doctor's identity. Hence, registration is required to set up accounts for doctors and link their identities with IC cards.

\par
The doctor's account should contain detailed profile information and the master key $Key_3$ extracted from the doctor's IC card through the One-time programming (OTP) interface. Aside from this, a corresponding identification number $ID_P$ and password are also configured for login through the programmer. Based on the profile information, HAS can check doctors' privilege of accessing a certain patient's IMD for carrying out dynamic access control. For example, all doctors that are in charge of a certain patient have the permission to export monitoring data from the patient's ICD, but only the chief physician can reconfigure it.

\subsubsection{Patient's Registration}
Similar to the doctors' registration, patients also need to set up accounts on the HAS which contain their profile information, the initial value ($IV$) and the master key $Key_2$.

\par 
Before the implantation, the doctor extracts the master key $Key_1$ from a new IC card through the One-time programming (OTP) interface and then fusing the OTP to prevent further access. After that, two master keys and $IV$ are loaded into the IMD and the cycle counter $i$ is set to 1. Once the implantation is completed, the IC card is delivered to the patient.
\begin{figure*}[t]
	\centering
	\includegraphics[width=\textwidth]{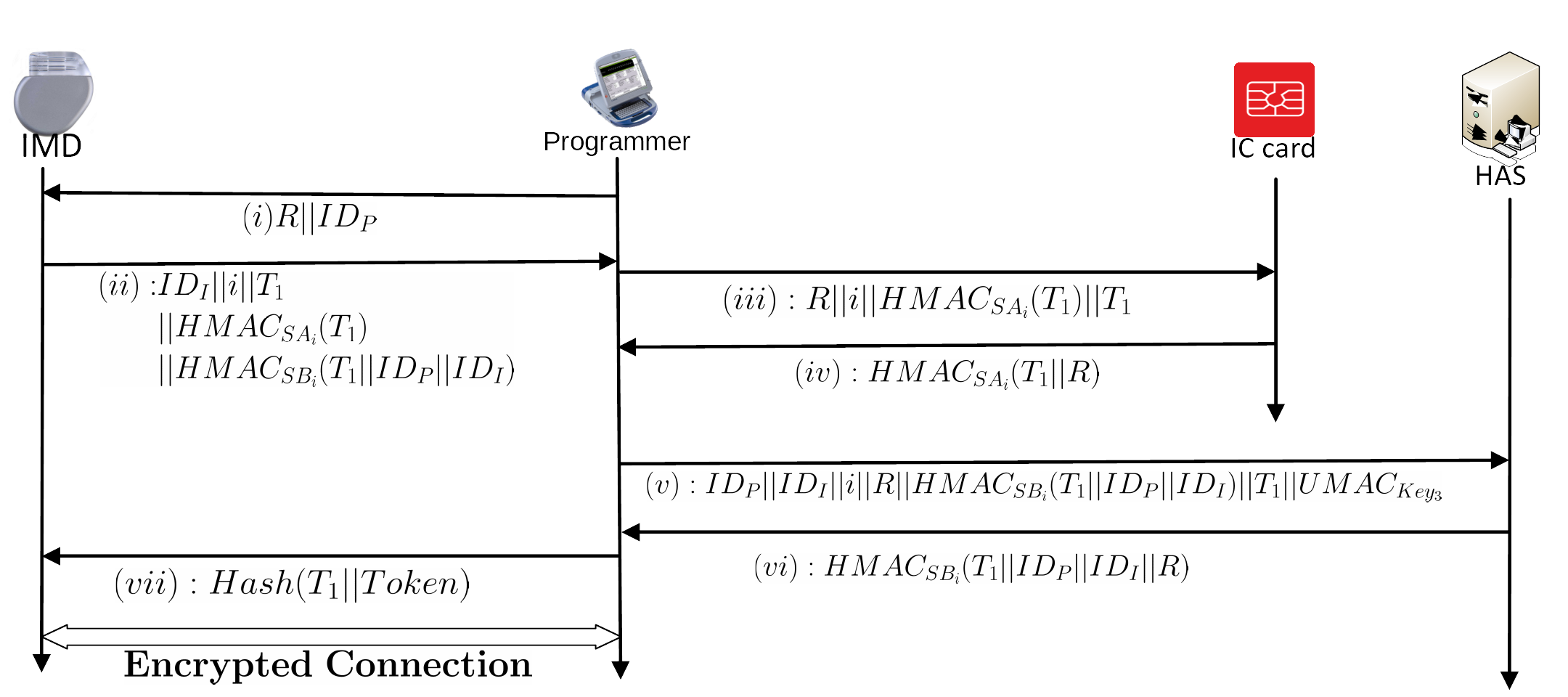}\\
	\caption{The authentication workflow.}\label{fig:flow}
\end{figure*}

\subsection{Service Request}
During normal access circumstances, a patient is conscious, and the wireless connection to the Internet is available. The doctor first inserts his/her IC card into the programmer and the programmer then logs in with his account name and password. Once the doctor's identity is successfully verified, a TLS session between the programmer and HAS is established to protect all communications between them, and the session is tagged by the doctor's identity number $ID_P$. Thereafter, with the patient's permission, the doctor has the patient’s IC card inserted into the programmer.

As shown by step $(\romannumeral1)$ in Figure~\ref{fig:flow}, the service request is initiated by the programmer sending the request code and the doctor's identity number to the IMD. When the IMD wakes up from periodical hibernation, it first retrieves the cycle counter $i$ and produce the current temporal key pair $SA_i$ and $SB_i$.  Then, a response containing the patient's identity number $ID_I$, counter $i$, current timestamp $T_1$, and two HMACs over these variables are sent back to the programmer (step $(\romannumeral2)$) as the challenge message.

\subsection{Authorization}
When the programmer receives a response from the IMD in step $(\romannumeral2)$, it forwards the two HMACs to the IC card and HAS respectively, as marked by step $(\romannumeral3)$ and $(\romannumeral5)$ in Figure~\ref{fig:flow}. Specifically, for message $(\romannumeral5)$, the programmer makes use of the doctor's IC card to generate a UMAC (a universal hashing based message authentication\cite{black1999umac}) for the entire message with the master key $Key_3$ as a proof of the doctor's identity.

\par
For the patient's IC card, it generate the temporary key $SA_i$ according to the received cycle counter $i$. With $SA_i$, it verifies the received HMAC on $T_1$. If the verification succeeds, the IC card sends back the HMAC on the timestamp $T_1$ cascaded by request code $R$ and increases its counter $i$ by 1.

Similarly, the HAS generates the temporary key $SB_i$  and verifies the HMAC of concatenated timestamp $T_i$ and two identities: $ID_I$ and $ID_P$. If the HMAC is valid, the HAS performs following checks:

\begin{enumerate}
	\item Check the temporal validity indicated by $T_1$ and time window length $TS$.
	\item Check the doctor's identity by verifying the UMAC with corresponding $Key_3$.
	\item Check the doctor's privilege level of accessing the the IMD with the identity $ID_I$.
\end{enumerate}

If all checks are successful, the HAS sends back the HMAC of a timestamp, two identity numbers and the request code $R$ as described by the step $(\romannumeral6)$ in Figure~\ref{fig:flow}. Finally, HAS's cycle counter $i$ is increased by 1.

\subsection{Session Establishment}
After the programmer receives the returned HMACs from the patient's IC card and the HAS, the access token is generated by performing XOR operation on the two HMACs as described in (\ref{equ:token}). Then the programmer sends the hash of timestamp $T_1$ and the token back to IMD (step (\romannumeral7)).

\begin{equation} \label{equ:token}
\begin{split}
token(i)=& HMAC_{SA_{i}}(T_1||R)\\
\oplus & HMAC_{SB_{i}}(T_1||ID_P||ID_I||R)
\end{split}
\end{equation}

On receiving the response from the programmer, the IMD first records the current time $T$ and checks whether the response is returned within the time window  $TS$. If the response is timely enough, the IMD generates the token $token'$  (as shown in (1)) by itself independently (the IMD possesses both two master keys to produce both two temporary keys $SA_i$ and $SB_i$). After that, the IMD compares the received $Hash(T_1||Token)$ with the generated $Hash(T_1||Token')$.  If two results match, the programmer is authenticated and authorized to access the IMD, and the cycle counter in the IMD is increased by 1. Otherwise, the IMD discards all completed steps and goes back to listening mode. Once the authentication and authorization is completed,  a secret number $SKey(i)$ is generated independently by the IMD and the programmer. The two parties calculate the XOR of two hashed HMACs as described in (\ref{equ:key}). The secret key  can be used as the seed for deriving session keys to encrypt the communication between the programmer and the IMD.

\begin{equation} \label{equ:key}
\begin{split}
SKey(i)= &Hash(HMAC_{SA_{i}}(T_1||R))\\
\oplus &Hash(HMAC_{SB_{i}}(T_1||ID_P||ID_I||R))
\end{split}
\end{equation}

\section{Enhancement}\label{sec:enh}
\subsection{Emergent Access}
Apart from the normal access control described in section \ref{sec:protocol}, special access mode is necessary for the emergent situation when the patient is unconscious and require emergency care. Under this situation, first aiders may need to access the patient's IMD to measurement the patient's physiological signal and perform timely treatment. However, the response message generated by the HAS may be inaccessible due to the lack of valid privilege or the absence of the Internet connections. To address this problem, we design an offline emergent access control scheme to use the temporal keys $SB_i$ cached inside the patient's IC card as an alternative of the HAS. For the security concern, the cached keys are obfuscated using the patient's iris code.

The HAS can generate patient specified number of temporal keys for future authentication rounds in advance and load them into the patient's IC card during the patient's normal visit to the medical center.  With the cached temporal key $SB_i$, the programmer can independently generate the correct response message as the step (\romannumeral6) in Figure~\ref{fig:flow}. When cached temporal keys are invalidated by regular accesses, the patient can easily the cached updtated by asking the HAS to generate more.

\subsection{Biometric Encryption of the Cached Temporary Key}

The drawback of the cached keys is the possible leakage of temporary keys. Since the cache keys are used as a substitution of HAS, disclosure of them would undermine the effectiveness of HAS. For instance, if attackers steal the patient's IC card, they can initiate access to the IMD immediately because both two temporary keys can be obtained from the IC card. Under this situation, the patient has no way to protect himself even he/she realizes the lost and report it to HAS manager. As a result, cached temporal keys must be encrypted to prevent unauthorized accesses. We denote each cache item as $$cache(i) = i||En_{C_k}(SB_i)$$ where $i$ is the counter and the $C_k$ is the cache encryption key. The cache encryption key $C_k$ is secured with the biometric encryption where the patient's iris code is exploited to obfuscate the original message as described in \cite{hao2006combining}.

Here, we give some brief retelling about the biometric encryption we use. First, a binary string is derived from an infrared image of patient's iris image by demodulating the phase information with complex-valued 2D Gabor wavelets \cite{daugman2004iris}. The generated reference string, denoted as $\Theta_{ref}$, is acquired in the patient register phase. After that, the HAS generates the cache encryption key $C_k$ and the key is encoded by Hadamard code and Reed-Solomon code. The encoded $C_k$ is then obfuscated by the iris reference $\Theta_{ref}$ and the resulted $\Theta_{lock}$ is stored in patient's IC card.

\begin{figure}
	\centering
	\includegraphics[width=0.6\textwidth]{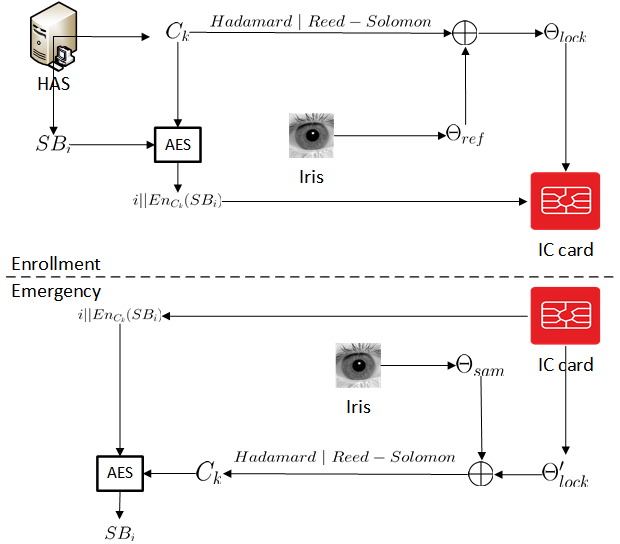}
	\caption{Iris code assisted emergent access.}
	\label{fig:emr}
	\vspace{-5pt}
\end{figure}

During emergent situation, first aiders can use a digital camera (with infrared mode) to capture the patient's iris image and generate the sample binary string $\Theta_{sam}$ to decrypt $\Theta_{lock}$ with the XOR operation and then acquire $C_k$ by doing Hadamard and Reed-Solomon decoding. After getting the XORed cache encryption key $C_k$, first aiders are capable of decrypting the temporary key $SB_i$ and access the patient's IMD locally without the HAS's support. The robustness of cache key regeneration is guaranteed by Hadamard and Reed-Solomon code which deals with errors in a binary level and block level respectively.

\subsection{Recovery Mode}
\begin{figure}
	\centering
	\includegraphics[width=0.6\textwidth]{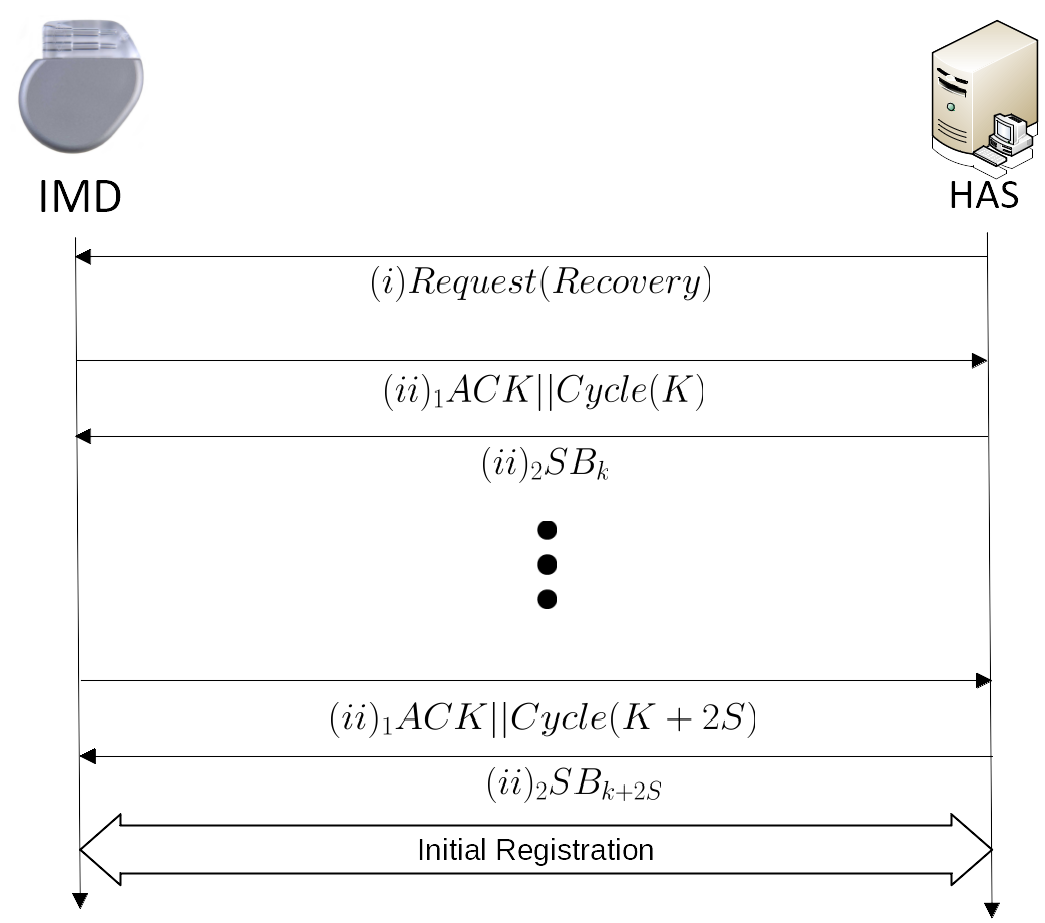}\\
	\caption{The recovery process.}\label{fig:repair}
\end{figure}

To deal with the issue of patients losing their IC cards, we design a recovery mechanism to reset the IMD and pair it with a new IC card. For the security concern of the recovery mode, the patient needs to go to the hospital to report the lost and sign the related documents in person before starting the reset procedure. A programmer operated by the security administrator is granted with the reset permission by the HAS. The recovery mode is triggered by the programmer with a designated service request code.  Upon receiving the reset request, the IMD begins to continuously challenge the programmer as shown in Figure~\ref{fig:repair}. During the challenge process, the IMD asks the programmer to provide the temporary key $SB$ for the cycle counter $K$. The challenge is conducted continuously for $2*S$ times where $S$ is the maximum number of cached temporary keys for the emergent access. The reset command is accepted and executed by the IMD if the programmer can provide correct responses to all those challenges with the help of the HAS. When the reset succeeds, the IMD rollback to the registration status to accept the configuration of new master keys $Key_1$ and $Key_2$. Thus, the doctor can re-do the patient's registration procedures to pair the IMD with a new IC card, and the old stolen IC card is invalidated permanently.

\section{Security Analysis}\label{sec:eva}
 
\subsection{Active Attacks}
In active attacks, adversaries intercept the transmission between the IMD and the programmer and then manipulate the message to deceive the them. However, the identities of the IMD, the programmer, and the HAS are secured by master keys or kept in the server or the POKs enabled IC card. Even if adversaries can successfully intercept those messages,  there is no way to retrieve the valid temporal keys for the current cycle. Without temporal keys, adversaries cannot generate valid HMAC for the tampered message. The reply attack is also not feasible due to the implementation of the time window. The IMD and the programmer would reject all duplicate packets that are outside of the time window.

\par
Another type of active attack is extracting credentials from the IC card through invasive ways. There have been a lot of methods to recover  security credentials by tampering IC card circuits\cite{fueki2001semiconductor,skorobogatov2010flash,chien2002stolen}. However,  with the master key secured by POKs, any attempt to tamper the circuit would fail and destroy the secret credential permanently.

\subsection{Desynchronization attack}
A lot of PUFs-based authentication protocols that rely on synchronization are subject to the desynchronization attack. Normally, the desynchronization attack is achieved by intentional interruptions during normal authentication, which results in the temporary key generator in different components running at malposed cycles. In our work, the desynchronization attack could be avoided by applying one-step cache in the IC card and the HAS for the previous used temporary keys $SA_{i-1}$ and $SB_{i-1}$. Since the IMD only increases its cycle counter $i$ when authorized by the IC card and the HAS succeeds, the desynchronization condition only happens when the IMD's counter is one step slower than the IC card or the HAS. When the IC card and the HAS receive the counter from the programmer at step $(\romannumeral3)$ and $(\romannumeral5)$ in Figure~\ref{fig:flow}, they can easily recover from the desynchronization status with the help of cached keys.

\subsection{Identity impersonation attack}
The adversaries may disguise themselves as legitimate physicians and use a programmer to access patients' IMDs. This kind of identity fraud is prevented in our scheme by the doctor's identity proof at step $(\romannumeral5)$ in Figure~\ref{fig:flow}. To impersonate a legitimate doctor, the attacker must get the corresponding doctor's identity IC card. Even if the attacker can steal a valid IC card from doctors, the stolen identities are not universally applicable to all IMDs, which restrict the range of damage. Also, this kind of attack cannot bypass the touch-to-access assumption because of requirement of the patient's IC card. It is difficult to use the IC card without the patient's permission. Through physical contacts with the patient, attackers are highly likely to be spotted and recorded by other people or video monitoring systems which greatly undermines the power of attacks.

\subsection{Security of the emergent access}
Many previous attempts introduced additional risk by adding an emergent access mode. For example, cloaker~\cite{gollakota2011they} and  IMDGuard~\cite{xu2011imdguard} rely on external devices to secure the IMD. When these external devices are removed, the IMD switches to open access. This can cause potential risk when the external device is stolen or the patient forgets to take it. Instead of trying to let the IMD be aware of the emergent situation, the IMD in our design always runs in the same mode, which means the aforementioned touch-to-access assumption always holds in emergent access since the temporary key $SA_i$ can only be acquired from the patient's IC card. To exploit the emergent access, attackers must get the patient's IC card and iris image. Even if the IC card is obtained by attackers without breaching the touch-to-access assumption (e.g., lost by the patient), the iris image must be captured in front of the patient. Additionally, cached keys in the IC card can be easily invalidated after the patient loses his/her card by rolling the normal access for multiple rounds to override the cached $SB$s. After that, attackers will not be able to access the patient's IMD even if they get the patient's iris code. The security of the iris code has been extensively discussed and many counterfeit iris detection schemes are proposed in \cite{wei2008counterfeit,kohli2016detecting,galbally2014image}. However, most of the counterfeit iris attack (e.g., cosmetic lens with texture, printed iris on paper) has the prerequisite of acquiring high resolution of the victim's Iris NIR image. The NIRs that are used to illuminate the iris only has the effective range of 50$\sim$70 cm \cite{Hei2011}. The successful iris code generation requires a resolution of at least 50 pixels in iris radius for the iris picture\cite{daugman2003importance}. As a result, it is hard for attackers to get the usable iris image without breaching the touch-to-access assumption.

\subsection{Security of Trivium}
Unlike the block cipher such as AES, there is not a widely recognized secure stream cipher. Trivium has been selected as part of the eSTREAM project~\cite{robshaw2008estream}. Up to now, no effective analysis attack is proposed that is better than the brute force search. Existing attacks utilizing the side channel like Differential Power attack~\cite{kazmi2017algebraic} and Fault Injection attack\cite{potestad2016fault} all require hardware tampering of the cipher IC. This kind of hardware attack is impractical in our design because of the use of tampering-resistant POKs. Once the circuit is tampered, the master key will be destroyed permanently. Moreover, the passive analysis is also not applicable to our design. The output of the Trivium stream generator, the temporary key $SA_i$, and $SB_i$ are never transmitted directly over the air. Instead, they are only used as keys for HMAC operations. 

\section{Evaluation}\label{sec:exp}
Because commercial IMDs are not opensourced for customization, we implement our design on our own testbed which comprises a TelosB sensor mote, a Raspberry Pi and a Laptop to simulate the IMD, programmer, and HAS respectively.  As stated in \cite{xu2011imdguard}, Telosb sensor mote is built on the similar low-energy platform as commercial IMDs making it a good choice to simulate the overhead of our system on real IMDs. With the prototype implementation, we evaluate the overhead of the consumption of energy, time, memory, and storage for both the computation process and communication.

\subsection{Overhead Statistics}
To evaluate the overhead, we make statistics about all computing and communicating operations conducted by the IMD during an access cycle as shown in Table~\ref{tab:comp_overhead} and Table~\ref{tab:comm_overhead}.

\begin{table}
	\centering
	\scriptsize
	\caption{Computing statistics.}
	\begin{tabular}{|c|c|c|c|}
		\hline
		
		Steps&Operation&Amount&Length (bits) \\ \hline
		Service Request&HMAC&$2$&$32,96$ \\ \hline
		Keypair Generation&Trivium&$2$&$256,256$\\ \cline{2-4}
		$ $&SHA-256&$2$&$256,256$\\ \hline
		Token Generation&HMAC & $2$ & $32,128$\\ \hline
		Session Key Generation&SHA-256&$2$&$256,256$\\ \hline
	\end{tabular}
	
	\label{tab:comp_overhead}
	\vspace{-5pt}
\end{table}

\begin{table}
	\centering
	\scriptsize
	\caption{Communication statistics.}
	\begin{tabular}{|c|c|c|c|}
		\hline
		
		Steps&Operation&Length (bits) \\ \hline
		Service Request&Receiving&$64$ \\ \cline{2-3}
		 & Sending & $608$	\\ \hline
		Token verifying&Receiving&$256$\\ \hline
	\end{tabular}
	\label{tab:comm_overhead}
	\vspace{-5pt}
\end{table}

\subsection{Experiment Results}  
We split the design into several parts, one for the key generator and others for the HMAC and SHA-256 operations. By deploying asynchronous counters, we record the time consumption of each part with the accuracy of 1 millisecond. With the energy consumption model \cite{prayati2010modeling,somov2009methodology,de2008energy} for TelosB as shown in Table \ref{tab:model}, the energy consumption for each operation is calculated from the time consumption.

\begin{table}
    \vspace{-5pt}
    \vspace{-5pt}
	\centering
	\scriptsize
	\caption{TelosB power consumption model.} \label{tab:model}
	\begin{tabular}{|c||c|c|}
		\hline
		
		$Operation$&$Power\ consumption (mW)$ \\ \hline
		$Transmit$&$69$\\ \hline
		$Listen$&$60$\\ \hline
		$Receiving$&$61$\\ \hline
		$Computing(active)$&$4.8$\\ \hline
		$Computing(idle)$&$4.5$\\ \hline
		$Sleep$&$0.035$\\ \hline
	\end{tabular}
	
	\vspace{-5pt}
\end{table}

The total computing time for a complete authentication cycle is  $367 ms$, if the transmission time is not taken into account. The overhead for different parts is listed in Table \ref{tab:overhead}.
\begin{table}
	\centering
	\scriptsize
	\caption{Overhead  for computation.}
	\begin{tabular}{|c||c|c|c|}
		\hline
		$ $ & $generator$ & $HMAC$ & $SHA256$\\  \hline
		$Time(ms)$ & $52$ & $46$ & $15$\\ \hline
		$Energy(\mu J)$ & $249$ & $220.8$ & $72$\\ \hline
		$ROM(bytes)$&$14290$&$6022$&$4792$\\ \hline
		$RAM(bytes)$& $312$&$282$&$197$\\ \hline
	\end{tabular}
	
	\label{tab:overhead}
\end{table}

Unlike the computational part, the overhead for transmission is much more complex because of the utilization of low power listening and collision avoiding mechanisms in the MAC layer. The listening time is uncertain depending on the environment noise, and the power consumption on synchronizations is difficult to specify.  Therefore, to make things easier,  we test the overhead of broadcasting packets with no retransmission. The results of the transmission test are listed in Table~\ref{tab:overheadt}. 
\begin{table}
	\centering
	\scriptsize
	\caption{Overhead  for communication.}
	\begin{tabular}{|c||c|c|}
		\hline
		
		$ $ & $Receiving$ & $Sending$\\  
		$ $&$(320\ bits)$&$(608\ bits)$\\ \hline
		$Time(ms)$&$40$&$22$\\ \hline
		$Energy(\mu J)$&$2440$&$1518$\\ \hline
		$ROM(bytes)$&$11140$&$11182$\\ \hline
		$RAM(bytes)$&$501$&$513$\\ \hline
		
	\end{tabular}
	
	\label{tab:overheadt}
	\vspace{-5pt}
\end{table}
		
With the operation specific evaluation results, ensemble time and energy overhead are calculate as illustrated in Equation (\ref{equ:energy}) and (\ref{equ:time}) .

\begin{equation}
\small
\centering
	\begin{split}\label{equ:energy}
	E =& E_{comp}+E_{Tx}+E_{Rx} \\
	=& E_{gen}+4*E_{HMAC}+3*E_{Sha256}+E_{Tx}+E_{Rx}\\
	=& 249\mu J+4*220.8\mu J+3*72\mu J+ 2440\mu J+1518\mu J\\
	=& 5306\mu J
	\end{split}
\end{equation}

\begin{equation}
\small
\centering
	\begin{split}\label{equ:time}
	T =& T_{comp}+T_{Tx}+T_{Rx} \\
	=& T_{gen}+4*T_{HMAC}+3*T_{Sha256}+T_{Tx}+T_{Rx}\\
	=& 52ms +4*46ms+3*15ms+40ms+22ms\\
	=& 343ms
	\end{split}
\end{equation}

The results indicate that our design introduces orders of magnitude lower energy and time overhead to the IMD compared with state-of-art physiological-feature-based solutions: The energy consumption of the OPFKA~\cite{OPFKA2013} ranges from about 70 $mJ$ to 1000 $mJ$ according to the coffer size. The IMDguard~\cite{xu2011imdguard} needs at least 45 seconds to measure more than 21 heart beats. While, the Heart-to-heart~\cite{CCS2013} uses TLS communication where the RSA encryption itself takes 5000000 cpu cycles (equivalent to 100 ms and 3038 $\mu J$ on the platform used by the author).

\section{Related Work}\label{sec:related}

\subsection{Pre-loaded-Key Based Solutions}
In some early works of IMD security, a long-term and device-specific credential is pre-loaded into the IMD and the programmer must possess the corresponding credential to pass the authentication. Halperin \emph{et al.} \cite{SP2008} propose a acoustic side channel based solution where programmer need a valid master key to access the IMD. Li \emph{et al.} \cite{HealthCom2011} present a rolling code based authentication scheme, in which the IMD and the programmer share a pair of encryption keys that are used to encrypt a sequence number. Liu \emph{et al.} Denning \emph{et al.} \cite{DenningCHI2010} uses visual objects to carry a static credential  for the IMD authentication. This type of IMD authentication/authorization schemes are obsoleted because the difficulty to protect the pre-loaded credential.

\subsection{Physiological Feature Based Solutions}
Recent researches tend to establish temporary keys by extracting time-varying information from patient's physiological signals. Most of them are also based on the touch-to-access assumption as our work does because the physiological features used for key generation can only be measured when the programmer is physically close to the patient. In \cite{CCS2013,xu2011imdguard,OPFKA2013,JBHI2016}, the patient's ECG features are utilized to derive credentials due to its high level of randomness. In \cite{Hei2011}, the access credentials are extracted from the patient's biometric features such as the fingerprints, iris and height.

This type of solutions face the problem of reliability: accurate measurement of physiological value is difficult because of the noise and distortion which results in long time consumption for feature measurement. Also, all these requires the IMDs to be equipped with special sensors to measure physiological signals such as ECG which not only increase the cost of IMDs but also causes excessive power consumption.

\subsection{Proxy Based Solutions}
Some researches propose to utilize additional devices to authenticate the programmer on behalf of the IMD for reducing the IMD's power consumption.The common design of proxy-based security schemes \cite{SIGCOMM,xu2011imdguard,Cloaker,ICC2014} includes (1) the proxy device jams the signal of external programmers until they are authenticated; (2) the IMD allows open-access when the proxy is not in proximity. 
		
The Shield \cite{SIGCOMM} jams the programmer's signal, but also the IMD's signal. Equipped with a full-duplex radio with two antennas, Shield is able to receive and decode the IMD's signal, meanwhile jamming it so that the programmer cannot receive and decode the complete IMD's message. A secured communication channel is assumed to have been set up between the Shield and the legitimate programmer. Therefore, Shield actually serves as a relay between the IMD and the programmer. Similarly, Cloaker \cite{Cloaker} is also a mediator that forwards all authorized communications between the IMD and programmer.

In the other schemes \cite{xu2011imdguard,ICC2014}, the proxy devices are mainly considered as an external authenticator, which stores the public keys of all authorized programmers and are able to verify the identity of the programmer that requests for access using its digital signature. The communication channel between the IMD and the programmer is established, after the programmer is successfully authenticated. In \cite{xu2011imdguard}, a pair of lightweight symmetric keys are issued to the IMD and the programmer, to encrypt their future communications. In \cite{ICC2014}, the authentication proxy is embedded into a gateway.

\par
This type of solutions requires patients to carry the active proxy device all the time which is unpractical. Also, the jamming may cause interference on other medical devices.

\section{Conclusion}\label{sec:con}
Security and low power consumption are among the most critical goals when designing access control schemes for implantable medical devices, while
how to attain both high energy efficiency and resiliency to powerful attacks, such as memory scanning and physical tampering, is still an unresolved challenge. We propose to take Physically Obfuscated Keys (POKs) as the hardware root of trust to establish a highly secure access control scheme for IMDs, and apply the
idea of computation offloading to saving energy consumption on the IMD side. We have comprehensively analyzed the security of the proposed scheme and compared it with some well-known mechanisms. In addition to its security advantage, a prototype system implemented on the TelosB platform demonstrates its high energy efficiency.

\section*{Acknowledgments}
This publication was made possible by NPRP grant \#8-408-2-172 from the Qatar National Research Fund (a member of Qatar Foundation). The statements made herein are solely the responsibility of the authors.

\bibliographystyle{splncs04}
\bibliography{cite}

\end{document}